# Observation of charge-to-spin conversion with giant efficiency at $Ni_{0.8}Fe_{0.2}/Bi_2WO_6$ interface

Saikat Das,[1,2*] Satoshi Sugimoto,[1] Varun Kumar Kushwaha,[1] Yusuke Kozuka,[1] and Shinya Kasai [1,3]

[1]National Institute for Materials Science, 1-2-1 Sengen, Tsukuba, 305-0047, Japan

[2] Department of Physics, Indian Institute of Technology Kharagpur, Kharagpur-721302, India

[3] Japan Science and Technology Agency, PRESTO, Kawaguchi, Saitama 332-0012, Japan

[*] Author to whom correspondence should be addressed: saikat@phy.iitkgp.ac.in

## Abstract

Magnetization switching using spin–orbit torque offers a promising route to developing non-volatile memory technologies. The prerequisite, however, is the charge-to-spin current conversion, which has been achieved traditionally by harnessing the spin–orbit interaction in heavy metals, topological insulators, and heterointerfaces hosting a high-mobility two-dimensional electron gas. Here, we report the observation of charge-to-spin current conversion at the interface between ferromagnetic $Ni_{0.8}Fe_{0.2}$ and ferroelectric $Bi_2WO_6$ thin films. The resulting spin–orbit torque consists of damping-like and field-like components, and the estimated efficiency amounts to about 0.48 ± 0.02, which translates to 0.96 ± 0.04 $nm^{-1}$ in terms of interfacial efficiency. These numbers are comparable to contemporary spintronic materials exhibiting giant spin–orbit torque efficiency. We suggest that the Rashba Edelstein effect underpins the charge-to-spin current conversion on the interface side of $Ni_{0.8}Fe_{0.2}$. Further, we provide an intuitive explanation for the giant efficiency in terms of the spin-orbit proximity effect, which is enabled by orbital hybridization between W and Ni (Fe) atoms across the interface. Our work highlights that Aurivillius compounds are a potential addition to the emerging transition metal oxide-based spin–orbit materials.

## I. Introduction

Electrical manipulation of magnetization via spin current-induced spin-orbit torque (SOT) has emerged as a promising pathway for developing next-generation spintronic memory and logic technologies.[1] Generating spin current requires utilizing spin-orbit coupling (SOC) to convert a charge current to its spin counterpart. Traditionally, the spin-Hall effect (SHE) in nonmagnetic heavy metal,[2] Rashba-Edelstein effect (REE) at inversion-asymmetric interface/surface,[3] and spin-momentum locked topological surface states[4] have been studied for charge-spin conversion. Under this scheme, the spin current emanating from the SOC host exerts a SOT on the magnetization of an adjacent ferromagnetic (FM) layer triggering a switching. Compared to the spin-transfer torque mechanism, the SOT-induced magnetization switching is faster and more energy efficient,[5] making the latter a topic of intensive research. One focus area concerns exploring new materials and strategies to enhance the charge-spin conversion efficiency or the SOT efficiency, which is defined as the ratio of spin current density to charge current density. Some of the approaches to enhance the SOT efficiency include the use of highly resistive $\beta$-phase W films,[6] asymmetric interfaces comprising of an ultrathin ferromagnet (Co) sandwiched between a heavy metal (Pt) and oxidized layer ($Al_2O_x$),[7] Fermi-level and interface-engineered topological insulator (TI)-based heterostructures.[8,9] In parallel, it is found that the oxidation of heavy metals like Pt and W dramatically improves the charge-spin conversion efficiency.[10,11] This observation signals a favorable prospect of oxides in developing highly efficient spintronic devices.

Transition metal oxides (TMOs) constitute a unique material class with a broad spectrum of functional properties like magnetism, ferroelectricity, and metal-insulator transition.[12] Such diverse behaviors originate from the complex interactions among charge, spin, orbital, and lattice degrees of freedom. The extreme sensitivity of these interactions to crystal symmetry and chemistry in TMO offers an unparalleled opportunity to control and engineer new functionalities, including SOC.[13] Regarding charge-spin conversion, the spin current yield from conducting Rashba $LaAlO_3/SrTiO_3$ interface is comparable to that of TI-based systems.[14] Meanwhile, a robust and symmetry-tunable SHE has been observed recently in 4*d* (5*d*) transition metal-derived oxide $SrRuO_3$ ($SrIrO_3$). The charge-spin conversion efficiency of these materials rivals that of elemental heavy metals.[15–17] In contrast, heterostructures consisting of an interface between insulating TMOs and metal have not been explored for charge-spin conversion.

In this work, using heterostructures consisting of ferromagnetic $Ni_{0.8}Fe_{0.2}$ (Py) and insulating (001)-oriented epitaxial $Bi_2WO_6$ (BWO) layers, we studied the possibility of charge-spin current conversion. BWO is a wide bandgap (≈ 2.7 eV) TMO with an orthorhombic layered structure that consists of alternating $Bi_2O_2$ sheets and pseudo-perovskite $WO_6$ blocks.[18] Oxidized Bi and W interfaces were previously found to yield charge-spin interconversion.[11,19] We, thus, posit that Py/BWO interface could also enable a charge-spin conversion. Another motivation to employ BWO is its ferroelectricity, with a characteristic Curie temperature (≈ 950 °C) and spontaneous polarization (≈ 50 μC $cm^{-2}$).[20,21] Thus, it provides a unique opportunity to investigate the scope of oxide ferroelectrics with strong SOC in spintronic applications.[22]

## II. EXPERIMENTAL METHODS

*Sample Fabrication*: All samples used in this work are grown with the pulsed laser deposition technique by employing the fourth harmonic (λ = 266 nm) excitation of an Nd: YAG laser on (001)-oriented $(LaAlO_3)_{0.3}$-$(Sr_2TaAlO_6)_{0.7}$ (LSAT) substrates. $Bi_2WO_6$ (BWO) layer was grown at 480–490 °C under an oxygen partial pressure of 120 mTorr while operating the laser at 15 Hz delivering energy of about 6 mJ/pulse.[23] After the growth of the $Bi_2WO_6$ layer, the samples were cooled to room temperature. Subsequently, $Ni_{0.8}Fe_{0.2}$ (Py) and $Al_2O_x$ layers were grown *in-situ* at a base pressure of 5 $\times 10^{-8}$ Torr by ablating Py and $Al_2O_3$ targets. The laser energy was set to 24 mJ/pulse and 16 mJ/pulse during the growth of Py and $Al_2O_x$ layers, respectively. The nominal BWO layer thickness is set to about 18 nm, while for the Py layer, the thickness $t_{Py}$ is set to about 5 nm. The Py layer thickness was controlled by counting the number of laser pulses, which was determined using a test sample that was measured using the X-ray reflectivity technique for the thickness calibration. After the growth, the combination of standard photo-lithography, Ar-ion beam etching, and lift-off techniques was employed to fabricate 20 × 100 $\mu m^2$ microstrips. Subsequently, electrodes comprised of Au (100 nm)/Ta (10 nm) were sputtered to realize devices for the spin-torque ferromagnetic resonance (ST-FMR) measurements.

*Structural Characterization*: The microstructure and energy-dispersive X-ray spectroscopy characterization were performed at room temperature using a transmission electron microscope (Titan G2 80-200, FEI). Meanwhile, the structural quality and topography were characterized by X-ray diffraction and atomic force microscopy techniques, respectively (Fig. S1, Supplementary material).

*ST-FMR measurements*: All the measurements reported in this article were carried out at room temperature and under ambient conditions. A signal generator (68369B, Anritsu) was used for supplying microwave current ($I_{RF}$) via the source (S) port, and the mixing signal was collected via the ground (G) ports of the ST-FMR device. The input signal was amplitude-modulated using an 8 kHz sinusoidal excitation of an amplitude of ~ 1 V from a lock-in amplifier (SR830, Stanford Research Systems). Subsequently, the DC mixing signal was then extracted using the lock-in technique. For DC-tuned ST-FMR measurements, a voltage source (GS200, Yokogawa Electric Co.) was used to sweep DC currents between −1.5 mA to 1.5 mA with a step of 0.3 mA. Measurements were repeated three times at each step to improve the signal-to-noise ratio, and their averaged response was analyzed. All measurements were carried out for an RF power of 6.31 mW.

## III. RESULTS AND DISCUSSION

We start by briefly highlighting the microstructure and chemical qualities of our sample. Figure 1a shows a schematic drawing of the $Al_2O_x$/Py/BWO/LSAT heterostructure, while in Fig. 1b, we show a representative high-angle annular darkfield scanning transmission electron microscopy (HAADF-STEM) image around the Py/BWO interface. It shows a layered structure on the BWO side, consistent with the (001)-oriented growth of the BWO layer (Fig. S1, Supplementary material). Meanwhile, the Py side of the interface features randomly orientated lattice fringes, suggesting the polycrystalline nature of the Py layer. Interestingly, the intensity of the HAADF image is noticeably suppressed near the interface, and we find this generic feature to be approximately 1 nm thick. To clarify the origin of this intensity suppression, we performed the energy-dispersive X-ray spectroscopy (EDS) measurement. Figure 1c shows the EDS maps highlighting the compositional distribution of Ni, Fe, W, and Bi atoms across the Py/BWO interface. The Ni-distribution exhibits a well-defined boundary, whereas the Fe-distribution gradually decays into the BWO side. In contrast, the dispersion of W and Bi atoms is limited within the BWO side, albeit with a noticeable heterogeneity near the interface. The EDS intensity profiles shown in Fig. 1d visually summarize these observations, where a reduction in the W and Bi intensity is accompanied by a sizable gain in Fe intensity (highlighted by triangles). These findings suggest that during the initial stage of Py deposition, high kinetic energy plasma species knock out W and Bi atoms from the topmost BWO layers and contribute to Fe diffusion from the growing Py layer.[24] Overall, these processes lead to Bi and W (within the BWO layer) and Fe (within the Py layer) deficiencies at the interface, which accounts for reduced HAADF intensity in Fig. 1b. Nonetheless, the above measurements confirm that the Py/BWO interface is of reasonable quality, with chemical disorders limited within the topmost BWO layer.

Next, we focus on the spin-torque ferromagnetic resonance (ST-FMR) measurements, which allow us to investigate a possible charge-spin current conversion in our sample. Figure 2a shows the circuitry and the example optical micrograph image of the device. During the measurement, we pass a radio-frequency (RF) charge current through the microstrip while applying an in-plane magnetic field ($H_{ext}$) at an angle $\theta = 35°$ from the current flow direction. If charge–spin current conversion occurs, the resulting spin current is expected to exert two distinct SOTs on the Py layer, namely, an in-plane anti-damping-like SOT ($\tau_{DL}$) and an out-plane field-like SOT ($\tau_{FL}$), as schematically shown in Fig. 2(a).[4] Consequently, the magnetization ($M$) of Py would undergoe an out-of-equilibrium precession, yielding an anisotropic magneto-resistive response. The magnetization ($M$) of Py consequently undergoes an out-of-equilibrium precession, yielding an anisotropic magneto-resistive response. The resulting oscillatory modulation in the resistance would then generate a rectified dc voltage ($V_{mix}$) through mixing with the RF-charge current, which can be detected using the lock-in technique while sweeping $H_{ext}$ to satisfy the FMR condition.

In Fig. 2b, we show the measured FMR spectra for excitation frequencies ($f$) in the range between 5–12 GHz. The resonance spectrum can be modeled with symmetric and antisymmetric Lorentzian functions.[2]

$$V_{\text{mix}} = V_{\text{S}} \frac{W^2}{(H_{\text{ext}} - H_{\text{o}})^2 + W^2} + V_{\text{A}} \frac{W(H_{\text{ext}} - H_{\text{o}})}{(H_{\text{ext}} - H_{\text{o}})^2 + W^2} \quad (1)$$

Here $W$, $H_{\text{o}}$, and $V_{\text{S}}(V_{\text{A}})$ are the spectral width, resonance field, and amplitude of the symmetric (antisymmetric) function, respectively. Figure 2c shows $V_{\text{mix}}$ spectrum around resonance (filled circles, $f$= 8 GHz), alongside the model fitting (black curve) with Equation (1), which allows estimating $W$, $H_{\text{o}}$, $V_{\text{S}}$, and $V_{\text{A}}$. Observation of a finite symmetric ($V_{\text{S}}$) and antisymmetric ($V_{\text{A}}$) components in the $V_{\text{mix}}$ signal, in conjunction with their characteristic angular and power dependence (Fig. S3, Supporting material), unambiguously confirms the existence of charge-spin current conversion and SOTs in the sample.[2] In this context, $V_{\text{S}}$ ($V_{\text{A}}$) is correlated to the in-plane (out-of-plane) SOT $\tau_{\text{DL}}(\tau_{\text{FL}})$, and follows the relation [10].

$$\frac{\tau_{\text{DL}}}{\tau_{\text{FL}}} = \frac{V_{\text{S}}}{V_{\text{A}}} \left(1 + \frac{4\pi M_{\text{eff}}}{H_{\text{o}}}\right)^{1/2} \quad (2)$$

Here $4\pi M_{\mathrm{eff}}$ refers to the effective magnetization and can be obtained by fitting resonance fields with the Kittel formula, $f = (\gamma/2\pi)[H_o(H_o + 4\pi M_{\mathrm{eff}})]^{1/2}$ [Fig. 2(d)]. Using the extracted $4\pi M_{\mathrm{eff}}$ = 5.12 ± 0.02 kOe and Equ. (2), the SOT ratio is evaluated at each $f$ and is shown in Fig. 2(e). The SOT ratio is fairly frequency-independent and adopts a mean value of 1 ± 0.1.

While the ST-FMR measurements confirm a charge-spin current conversion in the sample, they do not clarify whether it is of bulk or interfacial origin. Noteworthy, the BWO and $Al_2O_x$ layers are insulating in our sample. Thus, the charge-spin current conversion could only occur within the Py layer[25] or at the interface. However, the absence of a measurable ST-FMR signal on a controlled sample, $Al_2O_x$/Py /$Al_2O_x$ featuring symmetric top and bottom interfaces (see Fig. S4a in the Supplementary material) precludes the possibility of current conversion in the bulk Py layer and points towards an REE-induced interfacial origin.

Next, we consider the in-plane and out-of-plane SOTs and quantify the corresponding efficiencies, hereafter referred to as $\xi_{\mathrm{DL}}$ and $\xi_{\mathrm{FL}}$ respectively. To this end, first, we characterize $\xi_{\mathrm{DL}}$ using the DC-tuned ST-FMR technique;[26] subsequently evaluating $\xi_{\mathrm{FL}}$ from Equation (2), following the relation : $\tau_{\mathrm{DL}}/\tau_{\mathrm{FL}} = \xi_{\mathrm{DL}}/\ \xi_{\mathrm{FL}}$. In the DC-tuned ST-FMR technique, the RF charge current and a direct current (DC) are simultaneously applied through the microstrip. The resulting DC spin current and corresponding torque, $\tau_{\mathrm{DL}}$, thereby modify the FMR spectral width ($W$), which is related the magnetic damping parameter $\alpha_f$ $(= W\gamma/2\pi f)$, as following [2,26]

$$\alpha = \alpha_f + \frac{\sin\theta}{(H_o + 2\pi M_{\mathrm{eff}})\mu_0 M_{\mathrm{sat}} t_{eff}} \frac{\hbar \xi_{\mathrm{DL}} J_{\mathrm{DC}}}{2e} \tag{3}$$

Here, $\mu_0$, $M_{\mathrm{sat}}$, $t$, $\hbar$, and $e$ are the vacuum permeability, saturation magnetization, thickness of ferromagnetic Py layer, reduced Plank's constant, and electron charge, respectively. Meanwhile, $\xi_{\mathrm{DL}} = J_{\mathrm{S,DC}}/J_{\mathrm{DC}}$ relates the DC spin current density, $\hbar/2\mathrm{e}\, J_{\mathrm{S,DC}}$ to its charge counterpart, $J_{\mathrm{DC}}$. In Fig. 3, we plot the effective damping parameter $\alpha$ as a function $J_{\mathrm{DC}}$ for positive ($\theta = 35°$) and negative ($\theta = -145°$) magnetic field ($H_{\mathrm{ext}}$). $\alpha$ linearly scales with $J_{\mathrm{DC}}$, and a linear fit to data (solid lines) further demonstrates a sign reversal of the slope upon switching the $H_{\mathrm{ext}}$ polarity. These observations are consistent with the prevalent DC-tuned ST-FMR studies [2,26] and, therefore, reliably allow for quantifying $\xi_{\mathrm{DL}}$. Noting, $M_{\mathrm{sat}}$ =5.34×10$^5$ A m$^{-1}$, $t_{eff}$ = ($t_{\mathrm{Py}}$− $t_{\mathrm{d}}$) nm = 3.5 nm, where $t_{\mathrm{d}}$ =1.5 nm accounts for the interfacial dead layer thickness (see Fig. S5, Supplementary material), we estimated $\xi_{\rm DL}$ = 0.48 ± 0.02. Noting $\tau_{\rm DL}/\tau_{\rm FL} = \xi_{\rm DL}/\xi_{\rm FL} \approx 1$ [from Fig. 2 (e)] we estimated $\xi_{\rm FL}$= 0.48 ± 0.02. Similarly, we have studied another sample with $t_{\rm Py}$ = 10 nm, and found $\xi_{\rm DL}$ ($\xi_{\rm FL}$) = 0.62 ± 0.05 (0.41 ± 0.03).

Overall, the DC-tuned ST-FMR study reveals a robust SOT efficiency in our sample. However, while calculating $\xi_{\rm DL}$ and $\xi_{\rm FL}$, we assumed that the entire Py layer contributes to charge-spin conversion via REE. This assumption underestimates the intrinsic SOT efficiencies by a factor of $1/t_{\rm I}$, where $t_{\rm I}$ refers to the interfacial screening length that is relevant for the REE. Assuming $t_{\rm I}$= 0.5 nm for Py,[10] we estimated the interfacial SOT efficiencies to be extremely large, amounting to about = 0.96 ± 0.04 nm$^{-1}$. This value is comparable to those of contemporary spintronic materials, as shown in Fig. 4 (see Table S1, Supplementary material for a detailed comparison).

The above data and analysis unambiguously demonstrate that the REE enables a charge-spin current conversion at the interface of our sample, and the resulting spin current exerts in-plane and out-of-plane SOTs on the Py layer. While both the $Al_2O_x$/Py and Py/BWO interfaces can support the REE, typically a contribution of about 0.006% - 0.04% in $\xi_{\rm DL}$ is expected from the $Al_2O_x$/Py interface,[25] which is much smaller than what we observed in our sample. Hence we suggest that the contribution from the Py/BWO interface is dominant in our sample. This conjecture is qualitatively supported by the observation of a weak ST-FMR signal in a controlled sample with asymmetric interfaces: $Al_2O_x$/Py/Glass, which does not contain the BWO layer (see Fig. S4b in the Supplementary material). As a possible explanation for large SOT efficiency, we note that, at a metallic surface/interface, the REE-induced spin accumulation scales with the Rashba coefficient, $\alpha_{\rm R}$.[27] The Rashba coefficient depends on the strength of the interfacial electric field ($E_{\rm i}$) and spin-orbit coupling ($\zeta$ ) as $\alpha_{\rm R} \sim E_{\rm i}\,\zeta$. The interfacial electric field promotes the forbidden intersite (onsite) orbital mixtures that satisfy the selection rules, $\Delta l = 0$ and $\Delta m \neq 0$ ($\Delta l = \pm 1$ and $\Delta m = 0$).[28,29] The intersite orbital mixture enables conduction electrons to acquire a non-zero orbital momentum, which leads to the momentum-dependent band splitting via spin-orbit coupling.[31,32] In our samples, the work-function mismatch between Py and oxide layers determines the strength of $E_{\rm i}$. Also, we note that (001)-oriented BWO thin films exhibit in-plane ferroelectricity,[20] which precludes the polarization-bound charge-induced electric field parallel to $E_{\rm i}$. The work function of Py and $Al_2O_x$ is around 4.8 eV and 3.9 eV.[33]. Meanwhile, the work function of BWO is around 5.7-6.06 eV.[34] From these considerations, we conclude that $E_{\rm i}$ is comparable at the $Al_2O_x$/Py and Py/BWO interfaces, and a significantly enhanced effective spin-orbit coupling strength at the latter underpins its dominant contribution to the giant SOT efficiency.

Next, we consider hybridization between Ni/Fe and W orbitals across the Py/BWO interface to rationalize the enhanced effective spin-orbit coupling strength. In BWO, W 5*d* (O 2*p*) orbitals predominantly form the conduction band minima (valance band maxima);[18,35] while in Py, 4*s* and 4*p* orbitals form the conduction band, and ferromagnetism arises due to the exchange coupling among localized 3*d* electron spins via highly itinerant 4*s* and 4*p* electrons. Thus, when interface states are formed via hybridization among W 5*d* and Py orbitals, conduction electrons in Py ( with $\zeta_{Py}$ = 65 meV) experience a stronger atomic spin-orbit interaction around W ($\zeta_W$ = 367 meV) atoms.[32] The onsite *s*-*p* and *p*-*d* orbital mixing ($\Delta l = \pm 1$ and $\Delta m = 0$) thereby enhances the effective spin-orbit coupling strength, $\zeta$, for itinerant electrons at the Py/BWO interface. Considering $\alpha_R \sim E_i\, \zeta$, such spin-orbit proximity effect naturally then explains the giant charge-spin conversion or the SOT efficiency of the Py/BWO interface.

## IV. CONCLUSION

In conclusion, we have observed the Rashba-Edelstein effect-induced charge-spin current conversion at the interface between $Ni_{0.8}Fe_{0.2}$ and $Bi_2WO_6$ layers. Using the spin-torque ferromagnetic resonance technique, we demonstrated that the spin current exerts both an in-plane damping-like and out-of-plane field-like spin-orbit torques on the ferromagnetic layer. The calculated interfacial spin-orbit torque efficiencies amount to about 0.96 ± 0.04 $nm^{-1}$, comparable to those of contemporary spin-orbit materials that yield spin current through interfacial charge-spin conversion with giant efficiency.

This work introduces a new material class: Aurivillius oxides as potential candidates for charge-spin interconversion-based spintronics research. From the prospect of fundamental studies, several questions have to be addressed, such as the magnitude of the Rashba parameter, $\alpha_R$, and the microscopic nature of the spin-orbit proximity effect. Density functional theory calculation considering orbital hybridization between Auriviilius oxides and ferromagnet could shed light on these aspects.[36,37] A sizable number of Aurivillius oxides are ferroelectric at room temperature. Therefore, we anticipate that our results could serve as a reference for exploring electrically tunable Rashba spin-orbit interaction using ferroelectric oxides.[35] Regarding applications, Aurivillius ferroelectrics are Si-compatible, stable at BEOL processing temperature, and can be heterointerface with a range of nonmagnetic electrode materials.[38]

These attributes favorably hint at a niche application of oxide-based spintronic devices for room temperature operations.

## SUPPLEMENTARY MATERIAL

See supplementary material for X-ray diffraction, atomic-force microscopy, Oxygen EDS map, power and angle dependence of the ST-FMR signal, room temperature magnetization characterization, ST-FMR data on the controlled sample with symmetric and asymmetric interfaces, and a table summarizing detailed comparison of samples shown in Fig. 4.

## ACKNOWLEDGMENTS

We gratefully acknowledge Dr. Seiji Mitani for giving the access to the VSM and AFM setups for magnetization and topography characterization, respectively. We would like to acknowledge support from Dr.Atsuya Kurita for preparing STEM specimens. This work was supported by Grants-in-Aid for Scientific Research Grants No. 19H02604 and by JST FOREST Program, Grant Number JPMJFR203D (Y.K.).

## AUTHOR DECLARATIONS

### Conflicts of Interest

The authors declare no conflicts of interest

### Author Contributions

**Saikat Das**: Conceptualization (lead); Data curation (lead); Formal analysis (lead); Investigation (lead); Writing – original draft (lead); Writing – review & editing (equal). **Satoshi Sugimoto**: Investigation (supporting); Writing – review & editing (supporting). **Varun Kumar Kushwaha**: Investigation (supporting); Writing – review & editing (supporting). **Yusuke Kozuka**: Funding acquisition (lead); Supervision (equal); Validation (lead); Resources (equal); Investigation (supporting); Writing – review & editing (equal). **Shinya Kasai:** Supervision (equal). Resources (equal); Investigation (supporting); Writing – review & editing (supporting).

## DATA AVAILABILITY

The data that support the finding of this study are available from the corresponding authors upon reasonable request

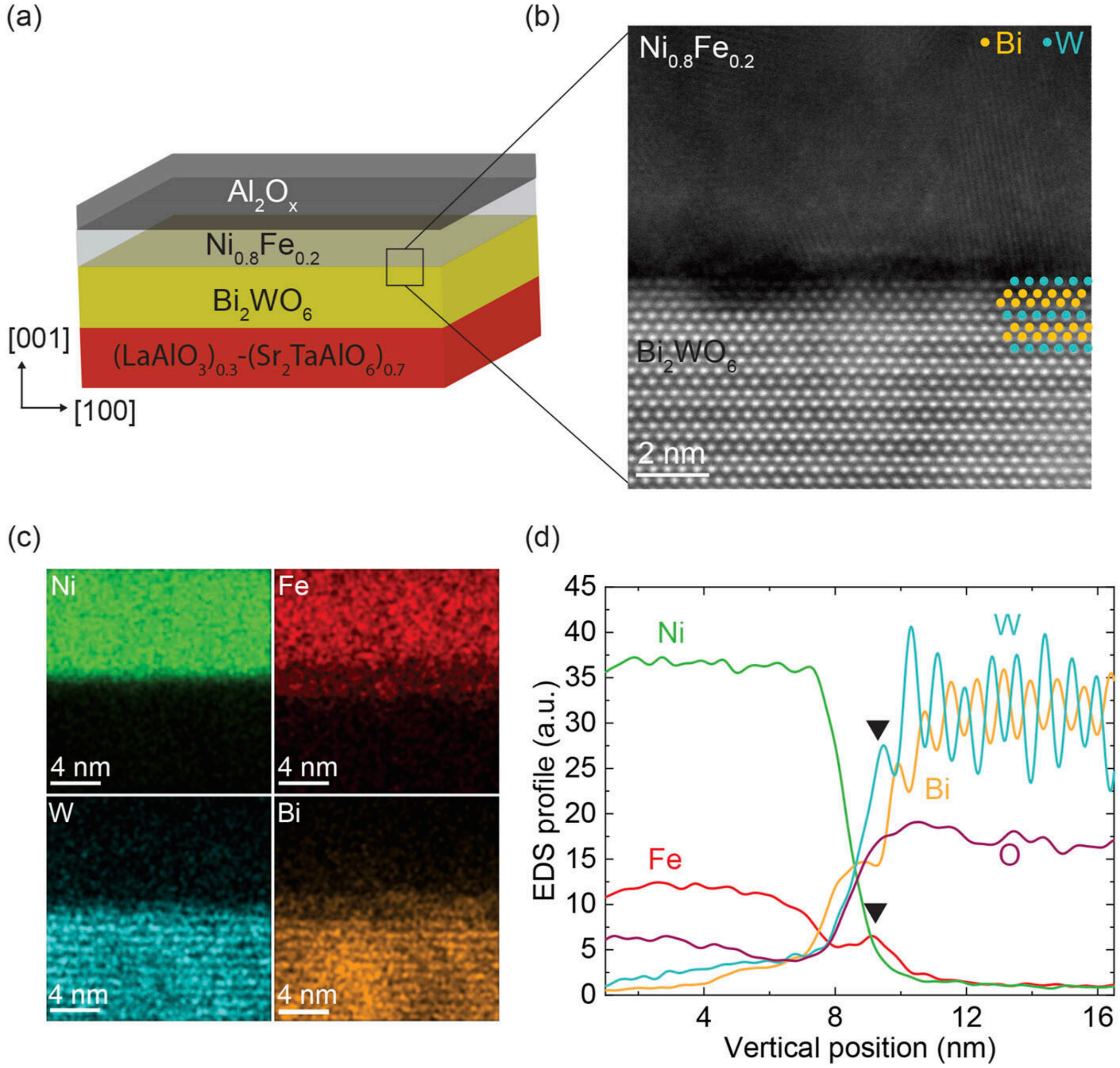

**FIG. 1.** Microstructure and elemental analysis. (a) Schematic drawing of $Al_2O_x/Ni_{0.8}Fe_{0.2}/Bi_2WO_6$ heterostructures. (b) High-magnification HAADF-STEM image around the $Ni_{0.8}Fe_{0.2}/Bi_2WO_6$ interface. We overlaid a schematic of Bi and W-layer sequence to highlight the (001)-oriented growth of the $Bi_2WO_6$ layer. (c) The EDS maps showcasing the elemental distribution around the $Ni_{0.8}Fe_{0.2}/Bi_2WO_6$ interface. The oxygen (O) EDS map is shown in Fig. S2 (supplementary material). (d) EDS intensity profiles obtained from the maps; black triangles highlight the accumulation of Fe atoms in the topmost layer of BWO and the Bi and W deficiency therein. The STEM images were obtained along the [010] zone axis of the cubic LSAT substrate.

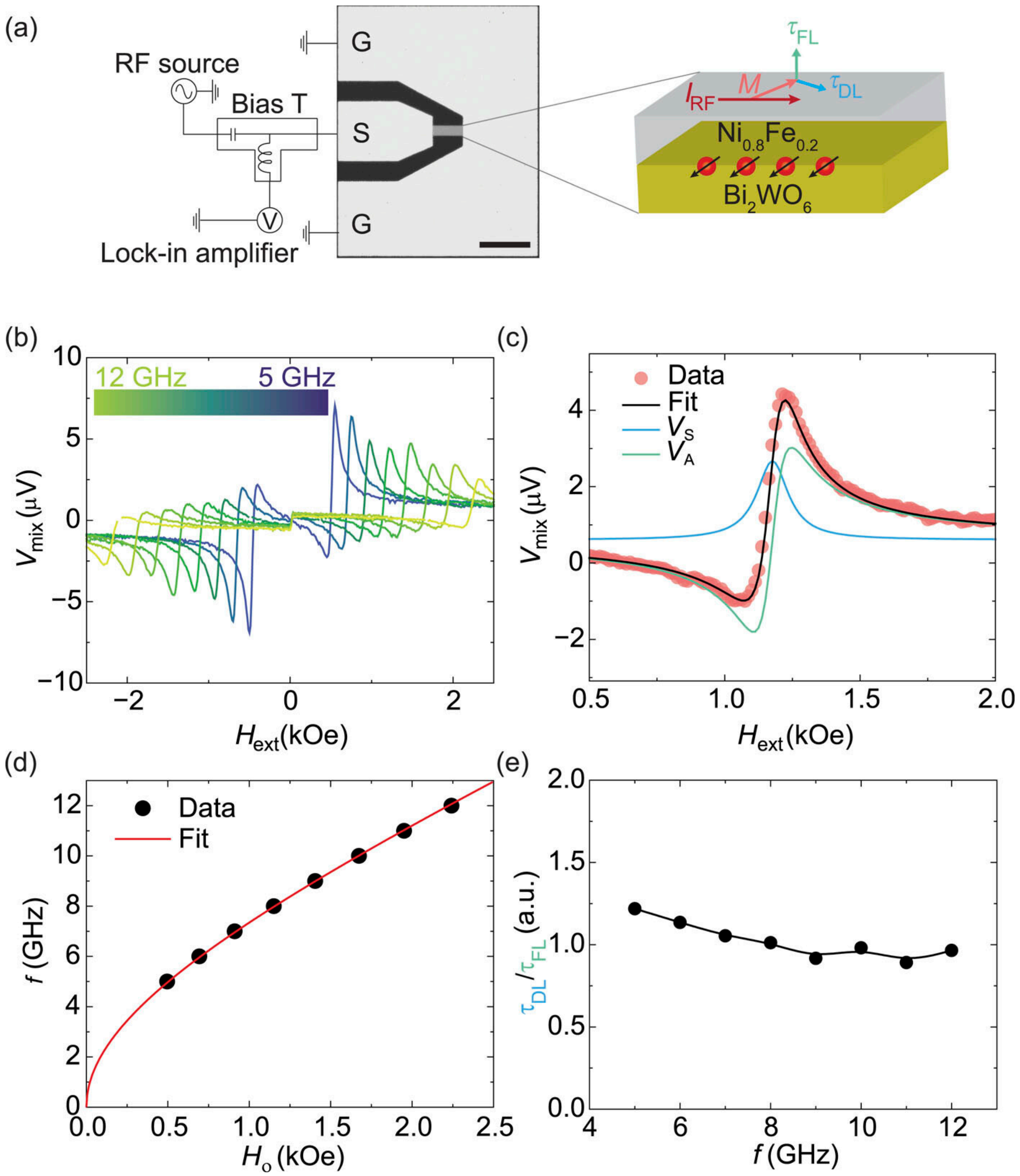

**FIG. 2.** Spin–torque ferromagnetic resonance (ST-FMR) measurement. (a) Optical micrograph of typical ST-FMR device and electrical circuitry used in our study. The inset schematically shows the geometric alignment of spin–orbit torque components relative to the magnetization ($M$) axis and charge cur- rent flow ($I_{RF}$). (b) ST-FMR spectra recorded by varying the excitation frequency, $f$, between 5 and 12 GHz. (c) Example fitting of an ST-FMR spectrum ($f$ = 8 GHz) with Eq. (1) around the resonance field. Solid blue and green curves correspond to the symmetric and antisymmetric Lorentzian components obtained from the best fit (black solid

line), respectively. (d) Fitting of the resonance fields, $H_o$ (black filled circles) according to Kittel's formula (red solid line) for extracting the effective magnetization. (e) Variation of spin–orbit torque ratio with $f$. The solid line is a guide for the eye.

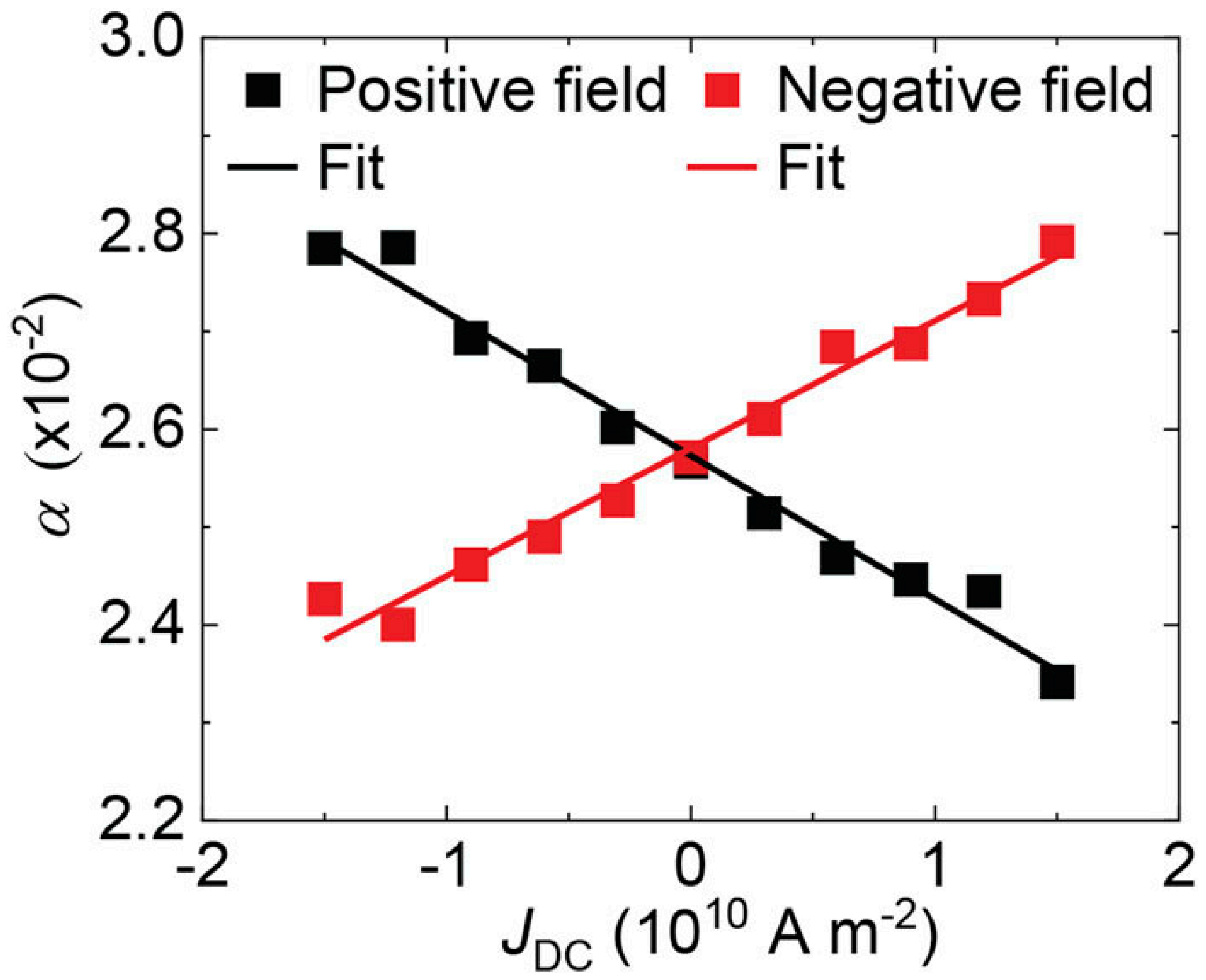

**FIG. 3.** Effective damping parameter (solid squares) as a function of DC charge current density. Solid lines represent the linear fits that are used to quantify $\xi_{DL}$ using Equation (3).

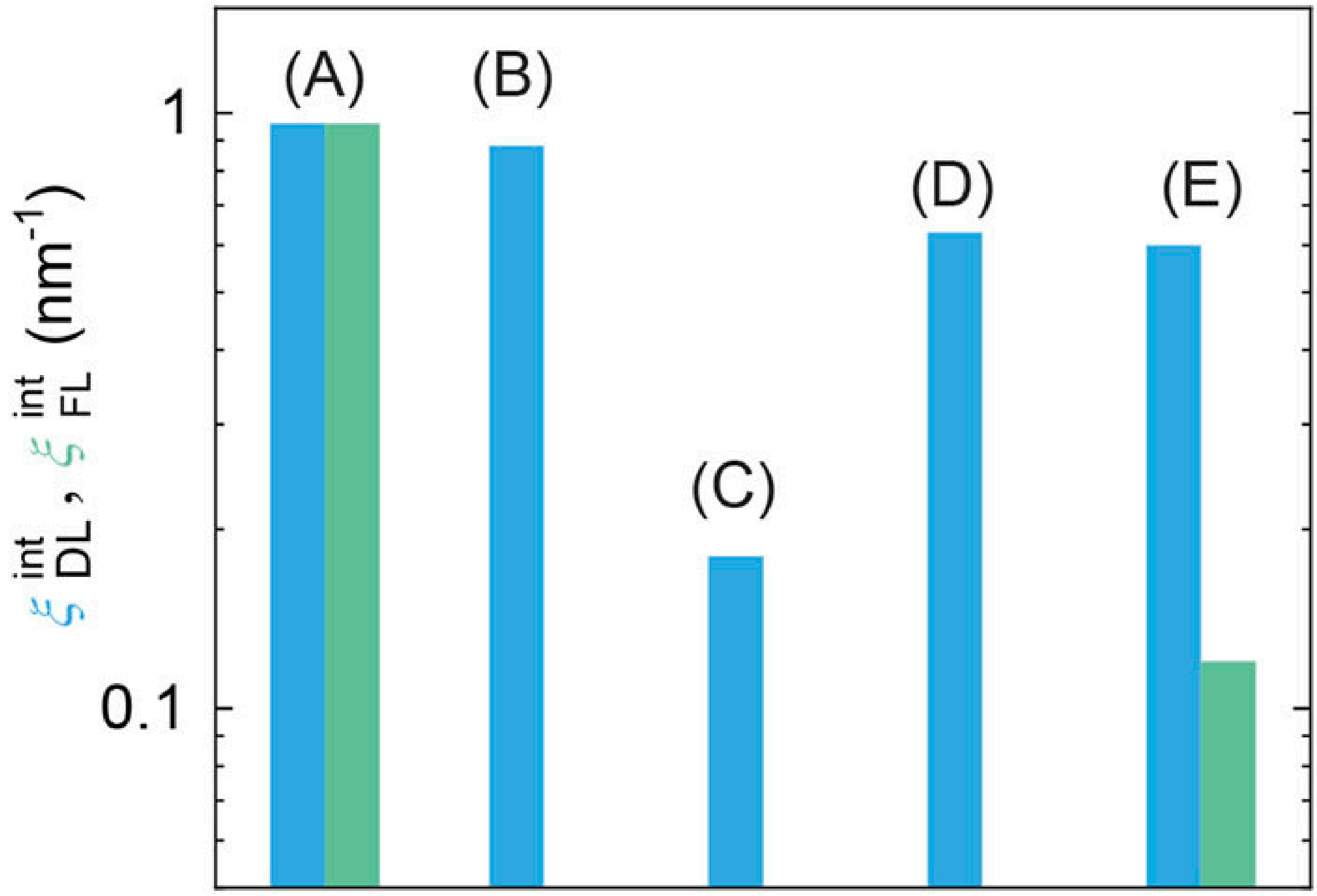

**FIG. 4.** Comparison of interfacial spin–orbit torque efficiencies of con- temporary spin–orbit materials. $Ni_{0.8}Fe_{0.2}/Bi_2WO_6$ used in this work (A), $Ni_{0.8}Fe_{0.2}/(Bi_{0.4}Se_{0.6})_2Te_3$(B),[8] $Ni_{0.8}Fe_{0.2}/LaAlO_3/SrTiO_3$ (C),[27] $CoFeB/LaAlO_3/SrTiO_3$ (D),[14] and $Ni_{0.8}Fe_{0.2}/PtO_x$ (E).[10] The nominal thickness of the ferromagnetic layer in samples (A) to (E) is about 5 nm.

# Supplementary Material:

**Observation of charge-to-spin conversion with giant efficiency at $Ni_{0.8}Fe_{0.2}/Bi_2WO_6$ interface**

Saikat Das,[1,2*] Satoshi Sugimoto,[1] Varun Kumar Kushwaha,[1] Yusuke Kozuka,[1] and Shinya Kasai [1,3]

[1]National Institute for Materials Science, 1-2-1 Sengen, Tsukuba, 305-0047, Japan

[2] Department of Physics, Indian Institute of Technology Kharagpur, Kharagpur-721302, India

[3] Japan Science and Technology Agency, PRESTO, Kawaguchi, Saitama 332-0012, Japan

[*] Author to whom correspondence should be addressed: saikat@phy.iitkgp.ac.in

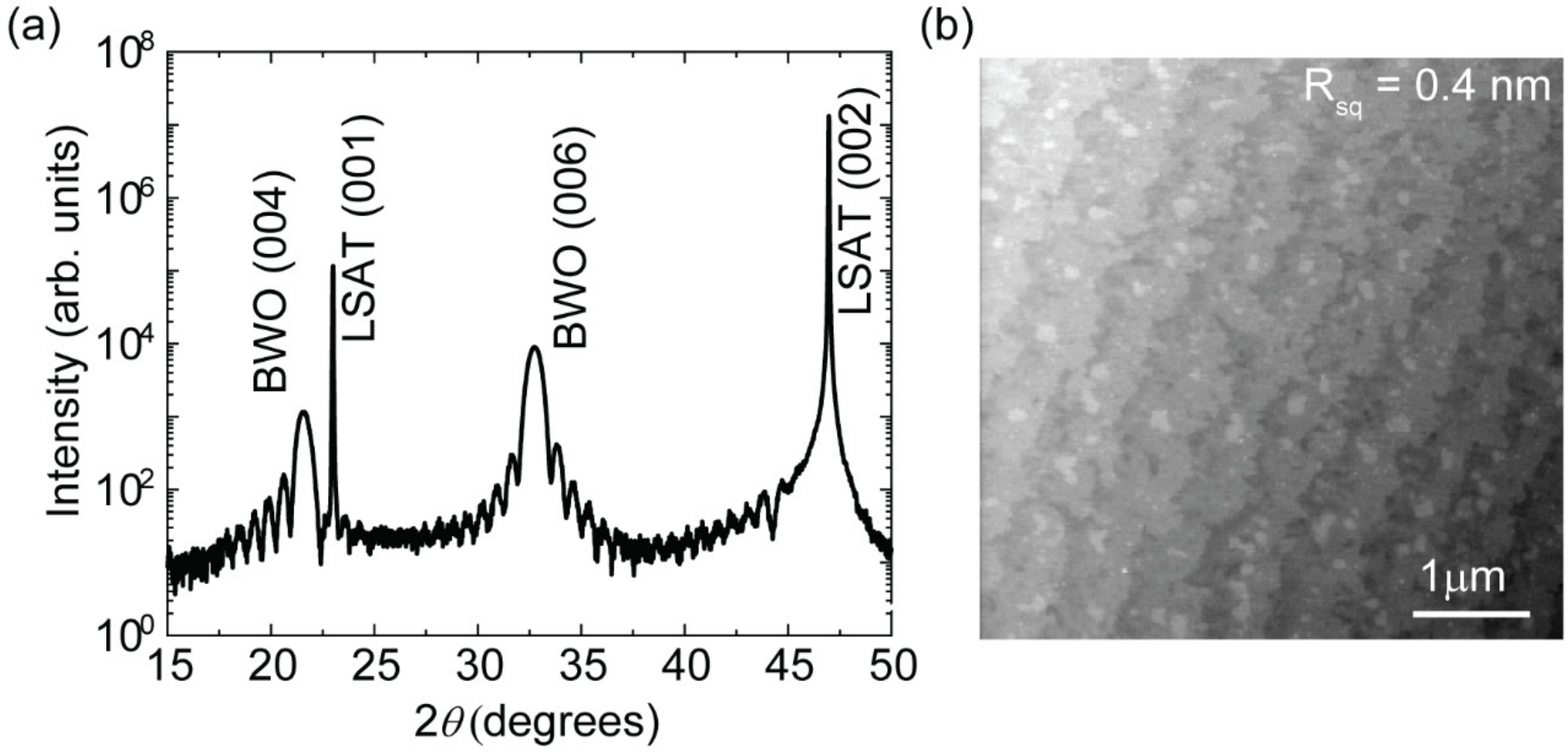

**Figure S1.** (a) Typical X-ray diffraction spectrum of epitaxial $Bi_2WO_6$ thin film grown on (001) oriented $(LaAlO_3)_{0.3}$-$(Sr_2TaAlO_6)_{0.7}$ (LSAT) substrate. (b) Example atomic force microscopy image a $Al_2O_x/Ni_{0.8}Fe_{0.2}/$ $Bi_2WO_6$ sample showing atomic step-terrace structure from the $Bi_2WO_6$ film. We estimated the root-mean-square roughness as $R_{sq}$ = 0.4 nm.

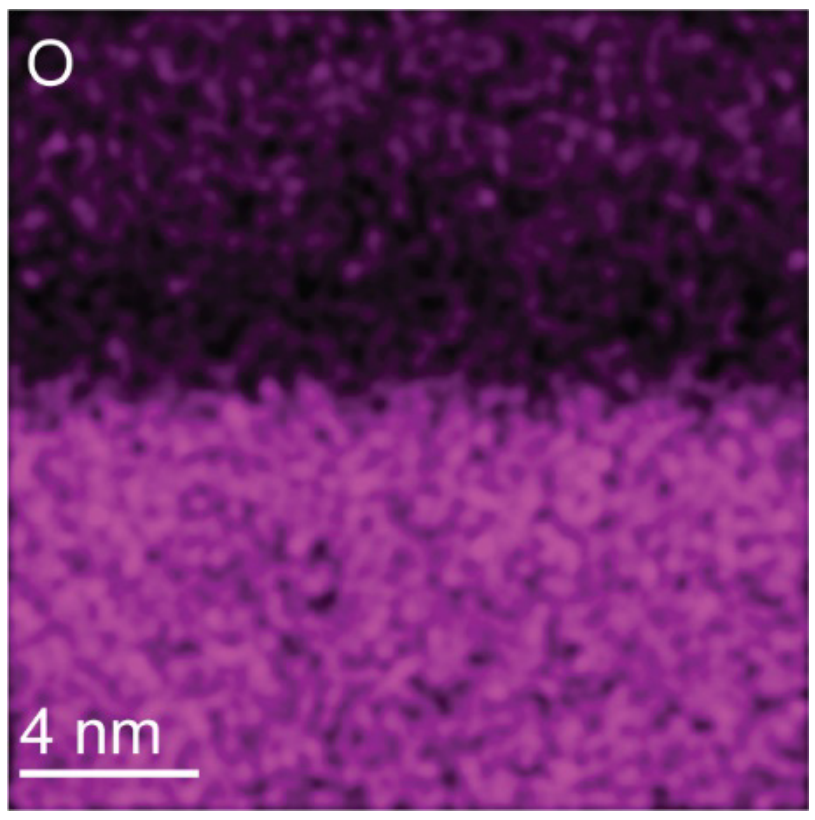

**Figure S2.** (a) Energy-dispersive X-ray spectroscopy (EDS) map of oxygen (O) across the $Ni_{0.8}Fe_{0.2}/Bi_2WO_6$ interface, which is captured along with the maps shown in Figure 1c of the main text.

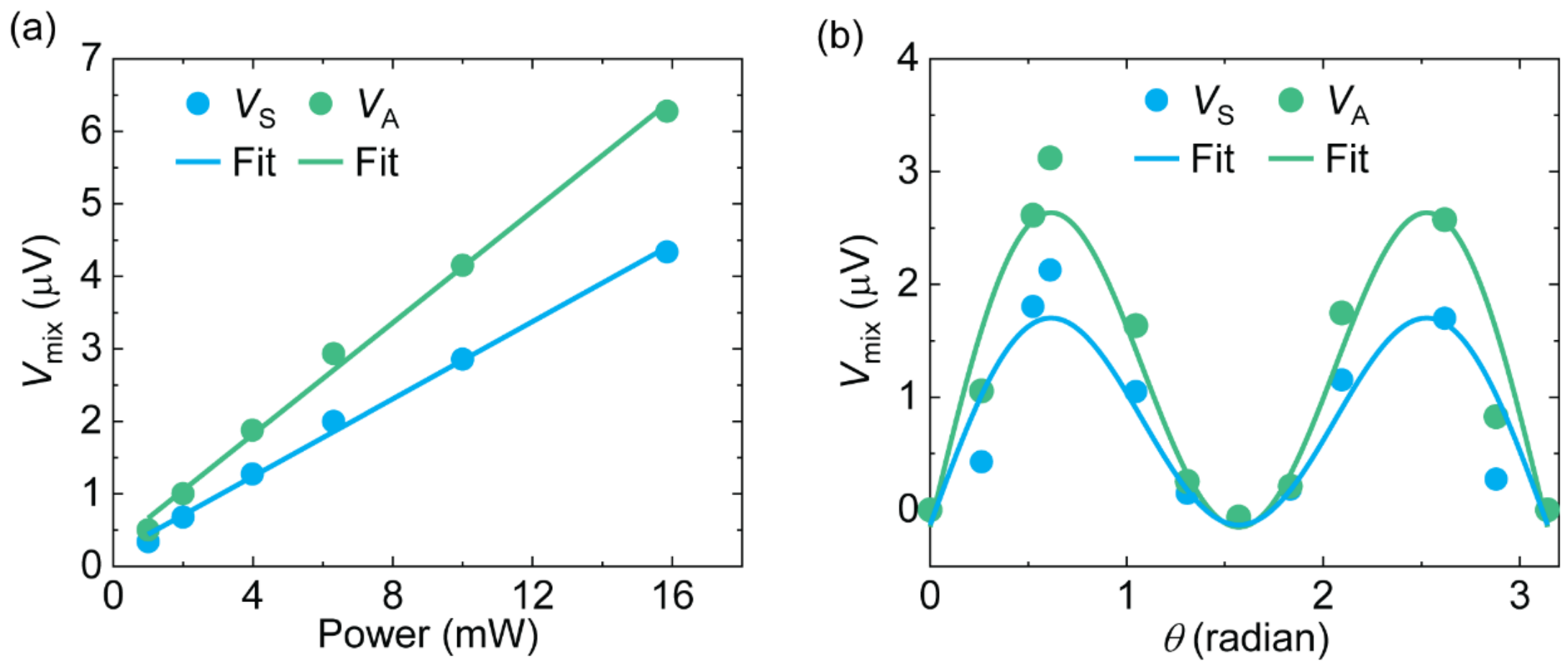

**Figure S3.** (a) Power dependence of the symmetric ($V_S$) and antisymmetric ($V_A$) components (solid circles) of rectified dc-voltage ($V_{mix}$) signals from our sample. The solid lines are linear fits to the data. (b) Variation of the symmetric ($V_S$) and antisymmetric ($V_A$) components (solid circles) of rectified dc-voltage ($V_{mix}$) signals. Here, $\theta$ refers to the angle between the RF current flow and the in-plane magnetic field applied during the ST-FMR measurement. The solid curves are fits by $\sim \sin 2\theta \cos\theta$.

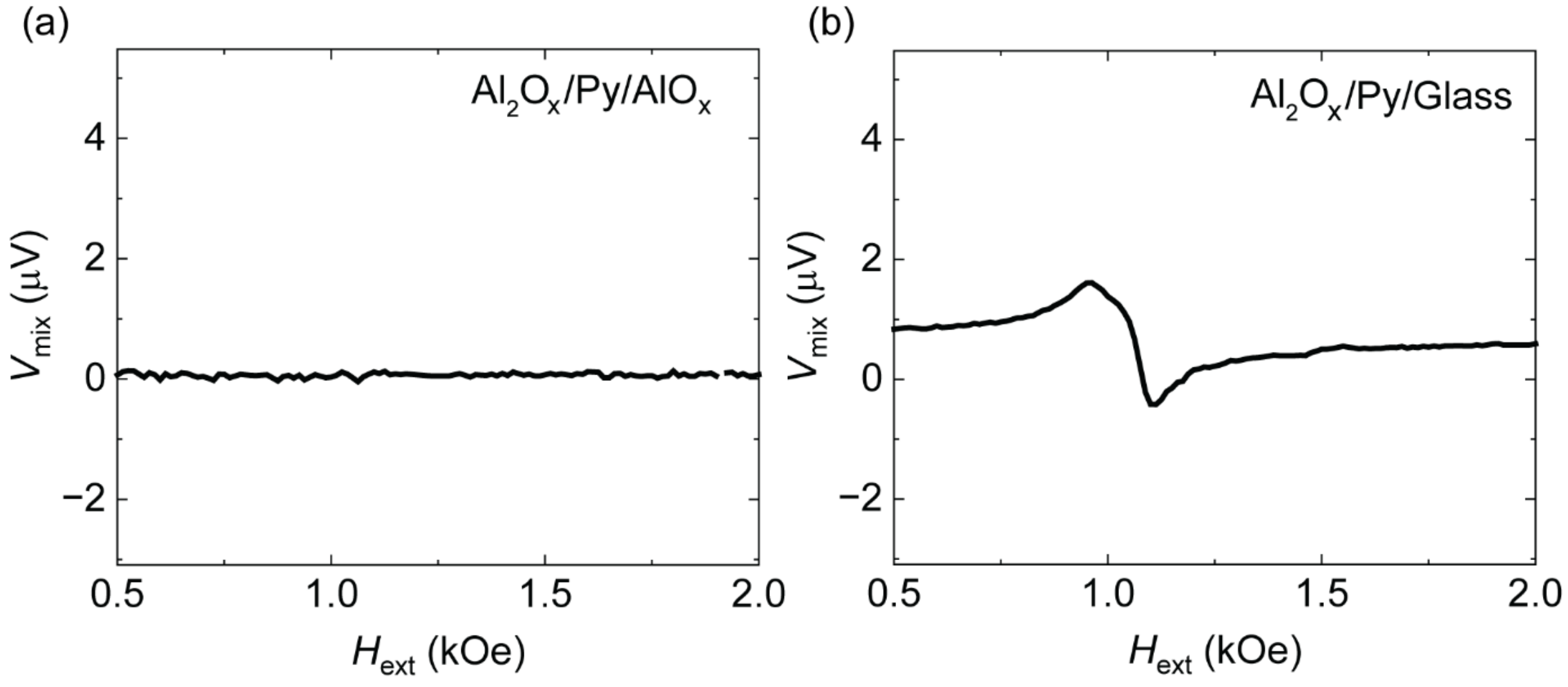

**Figure S4.** (a) ST-FMR measurement of a controlled sample with symmetric interfaces: $Al_2O_x$/Py/$Al_2O_x$. (b) ST-FMR measurement of a controlled sample with asymmetric interfaces: $Al_2O_x$/Py/Glass. Data shown here are collected at $f = 8$GHz.

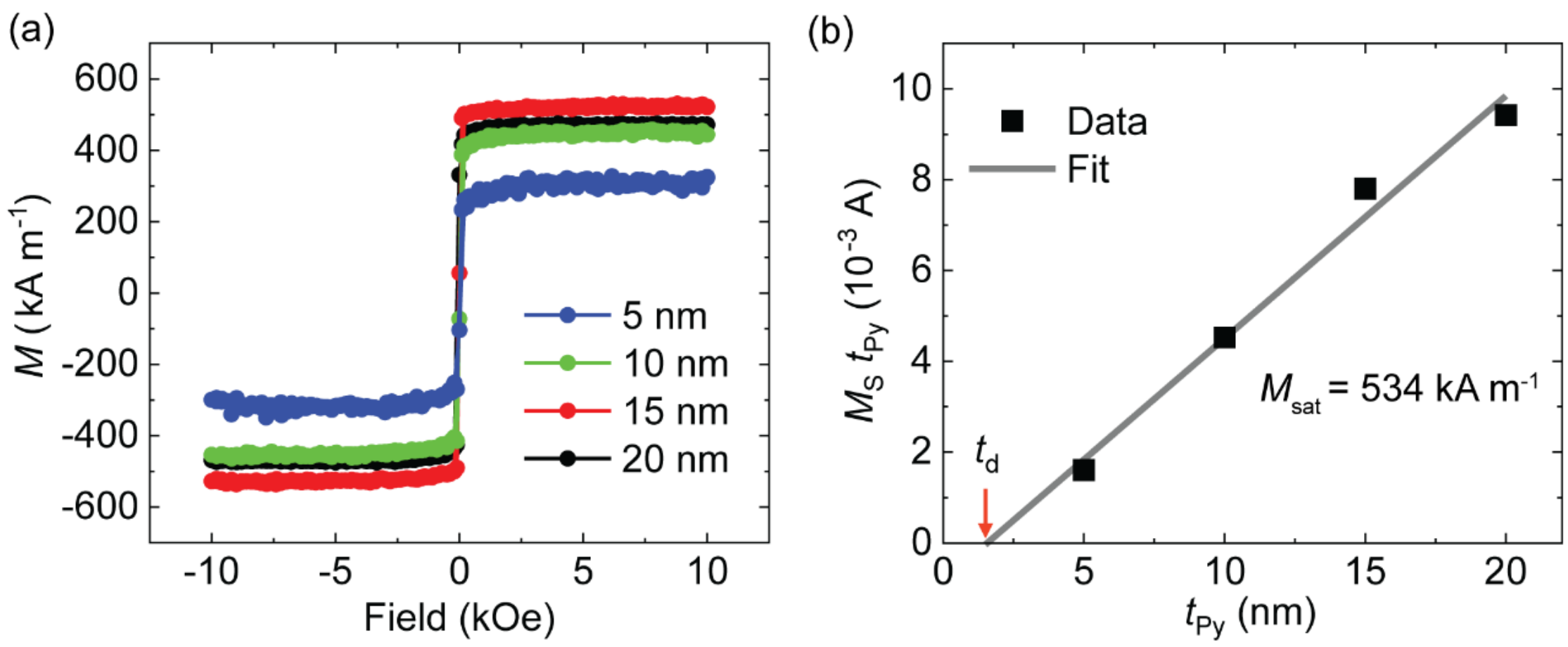

**Figure S5.** a) Room-temperature hysteresis loops of $Al_2O_x$/ $Ni_{0.8}Fe_{0.2}$/ $Bi_2WO_6$ samples. Here the thickness of the ferromagnetic layer is varied from 5-20 nm. b) Plot of the surface magnetization, $M_S t_{Py}$ (solid squares) as a function of $Ni_{0.8}Fe_{0.2}$ thickness, $t_{Py}$ . Here $M_S$ refers to the saturation magnetization extracted from Figure S6a. The solid line is a linear fit to data that highlights the presence of a magnetic dead layer in $Ni_{0.8}Fe_{0.2}$ (vertical arrow), and we estimated the corresponding thickness of $t_d$ = 1.5 nm. The slope corresponds to the intrinsic saturation magnetization $M_{sat}$ that we use to quantify the $\xi_{DL}$ using Equation (3) in the main text.

**Table S1:** Comparison of SOT efficiency between various samples that are used to plot Figure 4 in the main text. The nominal thickness of the ferromagnetic layer used in the following samples is about 5 nm.

| Samples | $\xi_{DL}$ | $t_I$ (nm) | $\xi_{DL}^{int}$ ($nm^{-1}$) | Method used |
|---|---|---|---|---|
| $Ni_{0.8}Fe_{0.2}/Bi_2WO_6$ | 0.48 | 0.5 | 0.96 | DC-tuned ST-FMR |
| $Ni_{0.8}Fe_{0.2}/(Bi_{0.4}Sb_{0.6})_2Te_3$ | 0.19-0.88 | 1 | 0.19-0.88 | ST-FMR, from $V_s/V_A$ ratio |
| $CoFeB/LaAlO_3/SrTiO_3$ | 6 | 10 | 0.6 | ST-FMR, $V_s$ only |
| $Ni_{0.8}Fe_{0.2}/LaAlO_3/SrTiO_3$ | 2 | 10 | 0.2 | ST-FMR, from $V_s/V_A$ ratio |
| $Ni_{0.8}Fe_{0.2}/PtO_x$ | 0.3 | 0.5 | 0.6 | DC-tuned ST-FMR |